\documentclass[letterpaper,twocolumn,prl,superscriptaddress,showpacs]{revtex4}
\pdfoutput=1

\usepackage{amsmath,amsfonts,amssymb}
\usepackage{graphicx}
\usepackage{enumerate}
\usepackage{bbm}

\newcommand{\ie}{\emph{i.e.}}
\newcommand{\se}[1]{\ensuremath{^{\text{#1}}}}
\newcommand{\si}[1]{\ensuremath{_{\text{#1}}}}
\newcommand{\vect}[1]{\ensuremath{\vec{#1}}}
\newcommand{\tens}[1]{\ensuremath{\boldsymbol{#1}}}
\newcommand{\T}{\ensuremath{^{\mathrm{T}}}}
\newcommand{\1}{\ensuremath{\mathbbm{1}}}
\newcommand{\orderN}{\ensuremath{\mathcal{O}(N)}}

\DeclareMathOperator{\tr}{Tr}

\newcommand{\ep}{\Phi}			
\newcommand{\epDL}{\Theta}		
\newcommand{\CepDL}{\vartheta}	
\newcommand{\inhom}{\gamma}	
\newcommand{\Vpp}{\Psi}		
\newcommand{\td}{\eta}			
\newcommand{\stm}{Q}			

\begin{document}
 
\title{Optimal Surface-Electrode Trap Lattices for Quantum Simulation with Trapped Ions}
\date{\today}
\author{Roman Schmied}
\affiliation{Max Planck Institute of Quantum Optics, 85748 Garching, Germany}
\author{Janus H. Wesenberg}
\affiliation{Department of Materials, University of Oxford, Oxford OX1 3PH, England}
\author{Dietrich Leibfried}
\affiliation{National Institute of Standards and Technology, Boulder, CO 80305, U.S.A.}
 
\begin{abstract}
	Trapped ions offer long internal state (spin) coherence times and strong inter-particle interactions mediated by the Coulomb force. This makes them interesting candidates for quantum simulation of coupled lattices. To this end it is desirable to be able to trap ions in arbitrary conformations with precisely controlled local potentials. We provide a general method for optimizing periodic planar radio-frequency electrodes for generating ion trapping potentials with specified trap locations and curvatures above the electrode plane. A linear-programming algorithm guarantees globally optimal electrode shapes that require only a single radio-frequency voltage source for operation. The optimization method produces final electrode shapes that are smooth and exhibit low fragmentation. Such characteristics are desirable for practical fabrication of surface-electrode trap lattices.
\end{abstract}
 
\maketitle

Trapped ions are a promising system for quantum computation and quantum simulation~\cite{Porras2006b,Blatt2008}. For the latter, simulations of coupled lattices that are ubiquitous in condensed-matter systems are of particular interest, as a ``proving ground'' for establishing the viability of large-scale quantum simulations as well as for studying poorly understood systems in great detail.

Trapping of ions over micro-fabricated planar surface-electrode structures has been demonstrated~\cite{Seidelin2006}, and the lithographic manufacturing process is well suited for producing complex electrode arrays~\cite{Amini2008}. Further, proposed schemes for local control over ion interactions using magnetic fields~\cite{Chiaverini2008,Ospelkaus2008} instead of laser beams have the potential to greatly simplify experimental setups. However, a dense lattice of microtraps cannot be constructed by just placing individual microtrap electrode sets side by side, because their electric fields overlap significantly and distort their microtraps. A regular dense lattice of ions for a two-dimensional condensed-matter simulation should be located at a height $z$ as far from the electrode plane as possible, because heating effects induced by the electrode surfaces scale strongly with this distance (proportional to $z^{-2}$ for Johnson noise~\cite{Wineland1998} and to $z^{-4}$ for anomalous heating~\cite{Turchette2000}). Moreover, this distance is crucial for directing laser beams onto the ions to induce spin-dependent interactions. The aforementioned cross-talk effects become prominent as the ion--surface distance $z$ is increased to about half the inter-ion spacing $d$. In this regime the surface electrodes for individual traps must be replaced by complex electrode patterns to generate a desired microtrap lattice. We present an algorithm that directly produces the periodic electrode structure for a desired lattice of trapping sites.

We consider ions of mass $m$ and charge $q$ that are confined by a time-averaged ponderomotive potential created by an inhomogeneous radio-frequency (rf) electric field with amplitude $\vect{E}(\vect{r})=-\vect{\nabla}\ep(\vect{r})$ at angular frequency $\Omega\si{rf}$. Additional electrodes inducing static electric fields could be incorporated for greater flexibility of design, but we do not consider this possibility here. In the adiabatic approximation, which assumes that the motion of the ions is slow on the time scale $2 \pi/\Omega\si{rf}$, the ponderomotive pseudopotential is~\cite{Dehmelt1967,Leibfried2003,House2008}
\begin{equation}
	\Vpp(\vect{r})=\frac{q^2 \|\vect{E}(\vect{r})\|^2}{4m \Omega\si{rf}^2}.
\end{equation}
Our aim is to create a periodic pseudopotential lattice with minima at a set of desired trap positions $\vect{r}_j$. We would like to specify the curvature of the pseudopotential at these sites, and also require that the rf field $\vect{E}(\vect{r}_j)$ vanishes, so that the ions undergo no rf-induced micromotion~\cite{Leibfried2003}. At a field-free point $\vect{r}_j$, the pseudopotential curvature tensor is proportional to the square of the electric potential curvature tensor $\tens{\ep}^{(2)}(\vect{r}_j)=\partial_{\alpha}\partial_{\beta} \ep(\vect{r}_j)$,
\begin{equation}
	\label{eq:curvatures}
	\tens{\Vpp}^{(2)}(\vect{r}_j)=\partial_{\alpha}\partial_{\beta} \Vpp(\vect{r}_j)
	= \frac{q^2}{2m\Omega\si{rf}^2} \tens{\ep}^{(2)}(\vect{r}_j) \cdot \tens{\ep}^{(2)}(\vect{r}_j).
\end{equation}
Since $\nabla^2\ep(\vect{r})=\tr\tens{\ep}^{(2)}(\vect{r})=0$, the principal curvatures of the pseudopotential cannot be chosen independently.
Our algorithm requires specifying the electric potential curvature tensors $\tens{\ep}^{(2)}(\vect{r}_j)$, which can be determined up to their irrelevant sign from the corresponding $\tens{\Vpp}^{(2)}(\vect{r}_j)$ through Eq.~\eqref{eq:curvatures},
provided that $\Vpp$ is fully confining at $\vect{r}_j$.

We can quantify the strength of a microtrap independently of the distance to the electrode plane $z_j$ and the potential $U\si{rf}\cos(\Omega\si{rf} t)$ applied to the rf electrodes by~\cite{footnote1}
\begin{equation}
	\label{eq:kappadef}
	\kappa_j = | \det\tens{\ep}^{(2)}(\vect{r}_j) |^{1/3} \times (z_j^2/U\si{rf}).
\end{equation}
For typical experimental parameters $U\si{rf}=50$\,V, $\Omega\si{rf}=2\pi\times200$\,MHz, and $z_j=30$\,$\mu$m, the geometric mean of the three principal trapping frequencies for $^9$Be$^+$ ions is $\bar{\omega} = \kappa_j \times 2\pi \times 53$\,MHz. The values $\kappa_j$ depend solely on the geometry of the surface electrodes. Small dimensionless curvatures $\kappa_j$ can to some extent be compensated by lowering $\Omega\si{rf}$, subject to the Mathieu stability requirements~\cite{Leibfried2003}, or by increasing $U\si{rf}$. However, much can be gained by optimizing the electrode shapes such that the $\kappa_j$ are maximized for given constraints.
The trap \emph{depth}, \ie, the ponderomotive pseudopotential $\Vpp(\vect{r}_{\text{sp},j}) = \td_j \hat{\Vpp}_j$ of the lowest saddle point $\vect{r}_{\text{sp},j}$ between a microtrap and the boundaries, is not optimized in the present algorithm, but will be given in what follows in terms of the energy scale $\hat{\Vpp}_j=q^2 U\si{rf}^2/(4m\Omega\si{rf}^2 z_j^2)$. For the above typical parameters, $\hat{\Vpp}_j = 4.7$\,eV.

The resources we consider for realizing general surface electrode arrays for trapped ions are two types of periodic electrodes in the $x y$ plane, one being grounded or at a slowly varying control potential, and the other carrying the rf potential at a uniform amplitude and phase. These electrodes are assumed to completely cover a single plane~\cite{House2008,Wesenberg2008}. Our algorithm optimizes the shapes of these electrodes to achieve maximal curvatures $\kappa_j$ at a set of $M$ field-free microtraps per unit cell at positions $\{\vect{r}_j\}_{j=1}^M$ and with electric potential curvature tensors $\{\tens{\ep}^{(2)}(\vect{r}_j)=C U\si{rf}\tens{\Gamma}_j\}_{j=1}^M$. The $3\times3$ matrices $\tens{\Gamma}_j$ must be symmetric and traceless to fulfill the Laplace equation. The dimensionless curvatures of the various microtraps are $\kappa_j=|C|\times z_j^2 | \det \tens{\Gamma}_j |^{1/3}$. Maximizing the common scaling factor $|C|$ therefore simultaneously maximizes all the $\kappa_j$ while preserving the relative curvatures of the different microtraps.

We subdivide the unit cell of the desired Bravais lattice into $N$ small patches, and label $\CepDL_i=\ep_i/U\si{rf}$ the dimensionless electric potential amplitude of the $i\se{th}$ patch electrode, with $\CepDL_i=0$ ($\CepDL_i=1$) signifying a connection to ground (rf). The algorithm below produces the optimal set of binary values $\CepDL_i\in\{0,1\}$ $\forall i=1\ldots N$ with a negligible number of instances where intermediate values $0<\CepDL_i<1$, which would require additional rf voltage sources, occur. The radio-frequency electric potential $\ep(x,y,0)$ in the electrode plane is fully determined by the shape of the surface electrodes, \ie, by the $N$-component parameter vector $\vect{\CepDL}$. This Dirichlet boundary condition can be extended from the electrode plane into the third dimension using a Fourier transform~\cite{House2008,Wesenberg2008} since the electric potential satisfies the Laplace equation $\nabla^2\ep(\vect{r})=0$: each Fourier component of in-plane wavenumber $\vect{k}=(k_x,k_y)$ is damped exponentially as $\exp(i k_x x+i k_y y-\|\vect{k}\|z)$ away from the electrode plane. The typical size of the electrode patches gives a natural cutoff in this Fourier series. The resulting dimensionless potential $\epDL(\vect{\CepDL};\vect{r})=\sum_i \CepDL_i \epDL_i(\vect{r})$ is then constrained to have the desired structure of $M$ microtraps per unit cell: at the position of each field-free microtrap the electric field must vanish,
\begin{equation}
	\label{eq:cond1}
	\vect{\nabla}\epDL(\vect{\CepDL};\vect{r}_j)=\sum_i \CepDL_i \vect{\nabla}\epDL_i(\vect{r}_j)=0,
\end{equation}
giving $3M$ linear conditions on the coefficient vector $\vect{\CepDL}$. Further, the curvature tensors of the electric potential at the microtrap positions must match the desired trap curvatures,
\begin{equation}
	\label{eq:cond2}
	\partial_{\alpha}\partial_{\beta}\epDL(\vect{\CepDL};\vect{r}_j) =
	\tens{\epDL}^{(2)}(\vect{\CepDL};\vect{r}_j)=\sum_i\CepDL_i\tens{\epDL}_i^{(2)}(\vect{r}_j)=C \tens{\Gamma}_j,
\end{equation}
giving another $5M$ linear constraints on $\vect{\CepDL}$. In principle we could add other linear local constraints, such as higher order derivatives of the electric field for controlling anharmonicities, to further customize the result to the situation at hand; however, doing so might result in a decrease of the maximum achievable curvatures. The set of $8M$ linear constraints are denoted $\tens{A}\cdot\vect{\CepDL}=C\vect{b}$ in matrix representation. We define $\vect{\inhom}=\tens{A}^+\cdot\vect{b}$ as their unique inhomogeneous solution which is orthogonal to the null space of $\tens{A}$, computed from the Moore--Penrose pseudoinverse $\tens{A}^+$. After decomposing $\vect{\CepDL}=C\vect{\inhom}+\vect{\CepDL}'$ with $\vect{\CepDL}'\cdot\vect{\inhom}=0$, the electrode optimization problem is to
\begin{itemize}
	\item find the vector $\vect{\CepDL}$ satisfying $\tens{A}\cdot\vect{\CepDL}'=\tens{A}\tens{\inhom^{\perp}}\cdot\vect{\CepDL}=0$, with $\CepDL_i\in\{0,1\}$ $\forall i=1\ldots N$, and maximizing $|C|=|\vect{\CepDL}\cdot\vect{\inhom}|/\|\vect{\inhom}\|^2$,
\end{itemize}
where we have used $\tens{\inhom^{\perp}} = \1-\vect{\inhom}\vect{\inhom}\T/\|\vect{\inhom}\|^2$ as the perpendicular complement of $\vect{\inhom}$. Such integer linear programs are typically nondeterministic polynomial-time (NP) hard to optimize~\cite{Papadimitriou}. Fortunately, relaxing the integer constraints to $0\leq \CepDL_i \leq1$ $\forall i=1\ldots N$ yields a linear program \cite{MatousekGaertner} that gives \emph{globally} optimal results within \orderN\ time and memory. The reason that relaxed constraints are sufficient in practice is that the optimal solution for $\vect{\CepDL}$ contains at most $8M$ ``interior'' values, $0<\CepDL_i<1$, and consequently at least $N-8M$ patch potentials are ``railed'' at either $\CepDL_i=0$ or $\CepDL_i=1$ and satisfy the original binary constraint. This is because the optimal solution of a linear program is a ``basic'' solution; by construction all basic solutions fulfill this railing condition~\cite{MatousekGaertner}. As the resolution of the patch decomposition is increased to infinity, the combined area of the fixed number of un-railed patches becomes vanishingly small, and the optimal electrode converges to a solution of the integer linear program in the sense that $\ep(x,y,0)/U\si{rf} \in \{0,1\}$ for all $(x,y)$. In practice, rounding the interior values to 0 or 1 has very little effect on the properties of the resulting potential even for modest grid resolutions.

\begin{figure}
	\includegraphics[width=\columnwidth]{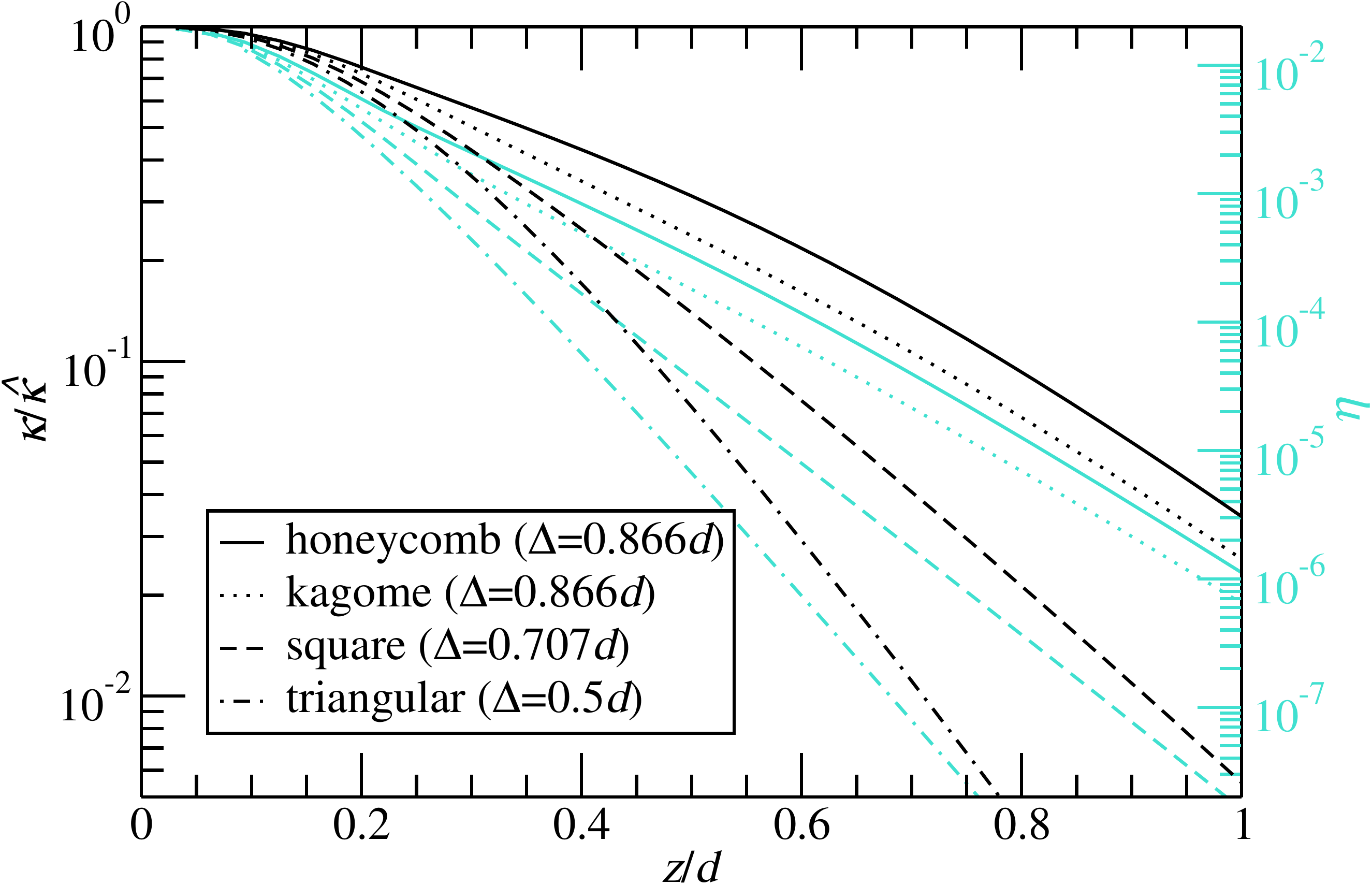}
	\caption{Optimized dimensionless curvatures $\kappa$ (black, left axis) and trap depths $\td$ (cyan, right axis) as functions of the ratio of microtrap height $z$ to inter-ion spacing $d$, for several lattice geometries. Curvature tensors are cylindrically symmetric out-of-plane quadrupoles as for the optimized ring trap ($\hat{\kappa}=0.298$ and $\hat{\td}=0.0196$; see text). The decreases in dimensionless curvatures and trap depths depend strongly on the typical electrode structure size $\Delta=2\pi/\stm$ (see text). Spurious microtraps may be present for some parameters.}
	\label{fig:curvatures}
\end{figure}

\begin{figure}
	\includegraphics[width=\columnwidth]{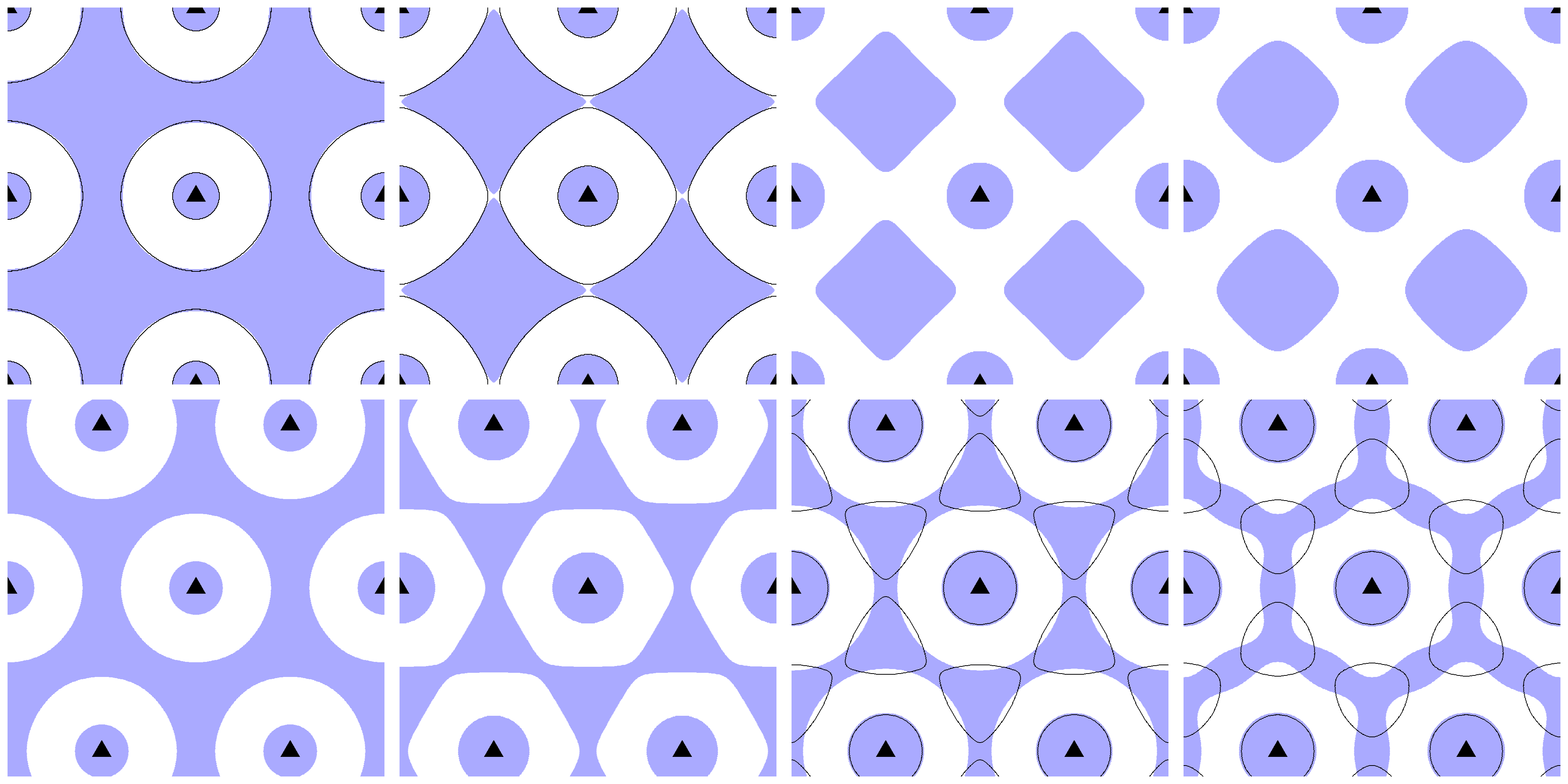}
	\caption{Top (bottom) row: optimized electrodes for square (triangular) lattices of microtraps at varying height $z/d=0.2$, 0.3, 0.4, 0.5 (0.25, 0.5, 0.75, 1) in units of the ion spacing $d$. Ground and rf electrodes are shown in white and blue (with arbitrary assignment). Microtrap locations are marked with triangles. Curvature tensors are cylindrically symmetric out-of-plane quadrupoles as for the optimized ring trap (see text). For $z/d = 0.2$, 0.3 (0.75, 1) the electrodes maximizing the dimensionless curvature $\kappa$ (outlines shown) give rise to spurious trapping sites which have been eliminated by imposing additional constraints (see text) reducing $\kappa$ by 0.07\,\% and 0.04\,\% (4\,\% and 18\,\%), respectively.}
	\label{fig:varyingheight}
\end{figure}

We have implemented the algorithm for oblique and hexagonal Bravais lattices. The oblique implementation uses $N=n_1\times n_2$ identical parallelogram-shaped patch electrodes per unit cell, while the hexagonal implementation takes $n_1=n_2=n$ and further divides each rhomb into two equilateral triangles, giving $N=2n^2$ patch electrodes. Optimizations with $N=10^6$ can be done in half an hour on a desktop computer. Our results are optimized with $n_1=n_2=n=250$, using $(2\times250)^2$ Fourier waves. We have found it unnecessary to use symmetries for specific wallpaper groups in the assignment of the patch electrodes, as the solutions found by interior-point methods of linear programming do not spontaneously break such symmetries, and the gain in computational time is only minor, due to the \orderN\ scaling.

The surface-electrode ion trap with the highest possible curvature is an rf ring electrode with a ratio of outer to inner radii of 4.98 embedded in an infinite grounded plane~\cite{Wesenberg2008}. We use its dimensionless curvature $\hat{\kappa}=0.298$ as a reference value. Regular sparse ($z/d\lesssim 1/2$) lattices of such cylindrical microtraps at a given trapping height $z$ are most easily constructed from lattices of similar ring electrodes, with the inner and outer radii re-optimized for each $z/d$ such that $\kappa$ is maximized. This simple parameterization already gives surprisingly good results, with values of $\kappa$ at most 10\,\% below the global optima, as determined by high-resolution linear programs (Fig.~\ref{fig:curvatures}). However, the outer radius must be smaller than $d/2$, which limits the possible trapping heights or necessitates more complex parameterizations. Nonparametric optimization, on the other hand, has solutions for any desired trapping height, as shown in Fig.~\ref{fig:curvatures}, albeit with exponentially decreasing dimensionless curvatures for large $z/d$ [$\kappa\propto (\stm z)^2 \exp(-\stm z)$ for $\stm z\gg 1$, with $\stm$ the wavenumber of the dominant Fourier component of the electrode pattern]. The dimensionless trap depths $\td$ are also found to decrease exponentially with $z/d$ [$\td\propto\exp(-2\stm z)$ for $\stm z\gg 1$]; however, the given values are not necessarily maximal. Figure~\ref{fig:varyingheight} shows that for low trapping heights the electrodes consist of almost circular ring electrodes centered under the desired microtrap locations; but at larger values of $z/d$ the electrodes of neighboring lattice sites overlap and interact significantly. This leads to nonintuitive patterns that, by construction, are optimal in the sense of maximizing $\kappa$.

\begin{figure}
	\includegraphics[width=\columnwidth]{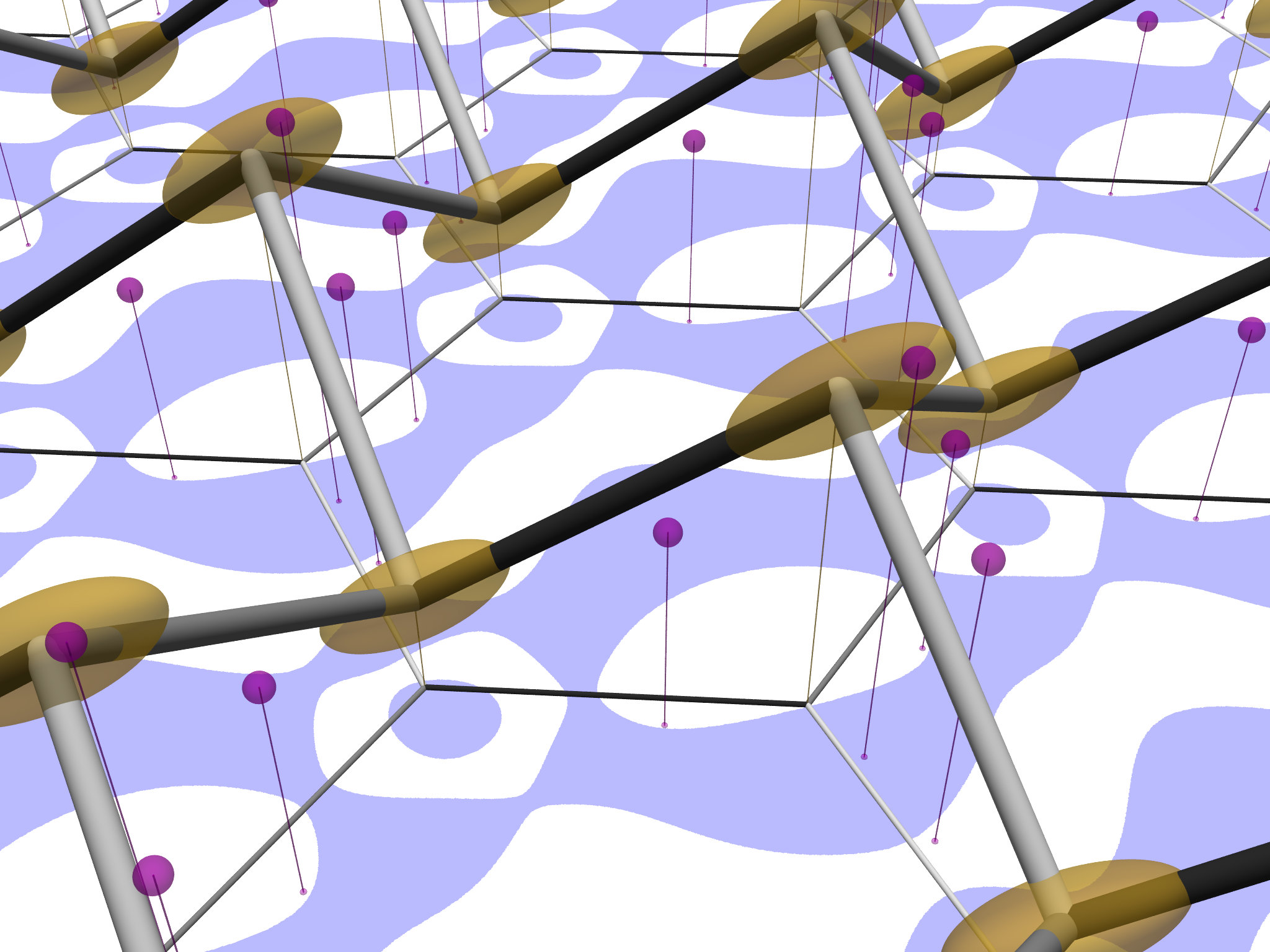}
	\caption{Optimized electrodes for a bilayer honeycomb lattice. The desired microtraps, shown as equipotential ellipsoids, are at heights $z_{\pm}/b=0.4\pm\frac{1}{2\sqrt{6}}$, with $b$ the honeycomb lattice periodicity and $d=b/\sqrt{2}$ the inter-ion distance. Vibrational frequencies of the microtraps are $(\omega_x,\omega_y,\omega_z)=\omega_0(\phi^{-1},1,\phi)$ in the three mutually orthogonal nearest-neighbor directions (marked with solid lines and projections; lighter shades indicate higher frequencies), where $\phi=(1+\sqrt{5})/2$ is the golden ratio. Dimensionless curvatures are $\kappa_-/\hat{\kappa}= 0.0022$ and $\kappa_+/\hat{\kappa}= 0.020$ (differing because of the different heights $z_{\pm}$). The locations of spurious microtraps are marked with small spheres. Thin vertical lines are added for clarity of perspective.}
	\label{fig:honeycomb2electrode}
\end{figure}

A major advantage of our nonparametric method is that much more complex trapping geometries can be constructed with no additional effort. However, for complicated microtrap structures the optimized dimensionless curvatures $\kappa_j$ and trap depths $\td_j$ tend to be vanishingly small, requiring unfeasibly large rf potentials for sufficient trapping. As an example, Fig.~\ref{fig:honeycomb2electrode} shows the optimized electrode for constructing a bilayer honeycomb lattice with right angles between nearest-neighbor directions~\cite{Micheli2006} and unequal trapping frequencies in these directions. The microtrap frequencies are chosen to be equal in the two trapping planes, leading to different dimensionless curvatures $\kappa_{\pm}$.

Figures~\ref{fig:varyingheight} and~\ref{fig:honeycomb2electrode} illustrate that optimal electrodes tend to consist of large smooth structures, with the un-railed patches located on the boundary between the rf and ground electrodes, reminiscent of anti-aliasing effects. This observation, which ensures that optimal electrodes can realistically be microfabricated, is attributed to the fact that microtraps can often be approximated by a localized cancellation of a small number of long-wavelength Fourier modes; the amplitudes of long-wavelength modes are maximized when the electrodes consist of large uniform areas, with sizes related to the wavelengths of the modes involved in the microtraps.

The linear programming algorithm is not limited to infinite periodic planar gapless electrodes. Any set of two- or three-dimensional electrodes can take the role of the ``patch'' electrodes, with an appropriate generalization of the method for extending the electric potential into the region where the ions are to be trapped. The algorithm will then specify which of these electrodes should be connected to rf and which to ground. In particular, \emph{finite} planar electrodes for a typical experimental setup may be optimized in order to avoid finite-size effects coming from the truncation of a periodic electrode array.

The algorithm offers unlimited freedom for placing arbitrary microtraps in arbitrary locations, and is optimal in the sense of globally maximizing the curvatures~\eqref{eq:kappadef} and producing smooth binary electrode shapes, but also has some limitations.
First, it offers no possibility to optimize the trapping \emph{depth} instead of its \emph{curvature}, because the former is a spatially nonlocal property and is therefore incompatible with the linear programming algorithm. Possibly parametric optimizations such as the ring-trap optimizations outlined above can be adapted to take the trap depth into account~\cite{House2008,Wesenberg2008}.
Second, specifying the desired set of microtraps does not imply that these will necessarily be the only microtraps generated by the optimized electrode set. Complex electrodes, such as those shown in Fig.~\ref{fig:honeycomb2electrode}, often generate a multitude of unwanted microtraps, predominantly on points or lines of symmetry of the underlying wallpaper symmetry group. In highly symmetric setups (Fig.~\ref{fig:varyingheight}), these spurious microtraps can often be eliminated by constraining the out-of-plane electric field $E_z(\vect{r})$ to judiciously chosen nonzero values at the spurious trap sites, thus adding further linear equalities or inequalities to the algorithm and reducing the optimized curvatures $\kappa_j$. In setups lacking symmetry (Fig.~\ref{fig:honeycomb2electrode}) this procedure is much less intuitive. In many cases it is possible to operate the trap in a regime where the spurious trapping sites are unstable, while the trapping sites at $\vect{r}_j$ are stable~\cite{Leibfried2003}.

The Coulomb interaction between ions at different trapping sites can lead to net static forces, due either to a finite lattice or to an insufficiently symmetric or non-planar unit cell (giving out-of-plane forces in Fig.~\ref{fig:honeycomb2electrode}). Such static forces tend to push the ions away from their rf zeros, thus inducing unwanted micromotion. Efforts should be made to compensate these forces by suitable control potentials. For this purpose we can subdivide the unit cell into control electrodes and ``patch'' electrodes \emph{before} optimizing the latter with our algorithm; alternatively, the \emph{optimized} ground and rf electrodes can be subdivided into areas biased with different control voltages. The same approaches are readily generalized to non-periodic ion trapping setups.

The authors thank Ignacio Cirac for initiating the collaboration, and Hermann Uys and Christian Ospelkaus for helpful comments on the manuscript. RS was supported by the European Union through the SCALA integrated project, and by the German Research Foundation (DFG) through the cluster of excellence \emph{Munich Center for Advanced Photonics} (MAP) and through program FOR~635.
JHW acknowledges the Carlsberg Foundation and the QIP IRC (Grant No.\ GR/S82176/01).
DL acknowledges support by IARPA and NIST.


\bibliography{MPQ}

\begin{thebibliography}{15}
\expandafter\ifx\csname natexlab\endcsname\relax\def\natexlab#1{#1}\fi
\expandafter\ifx\csname bibnamefont\endcsname\relax
  \def\bibnamefont#1{#1}\fi
\expandafter\ifx\csname bibfnamefont\endcsname\relax
  \def\bibfnamefont#1{#1}\fi
\expandafter\ifx\csname citenamefont\endcsname\relax
  \def\citenamefont#1{#1}\fi
\expandafter\ifx\csname url\endcsname\relax
  \def\url#1{\texttt{#1}}\fi
\expandafter\ifx\csname urlprefix\endcsname\relax\def\urlprefix{URL }\fi
\providecommand{\bibinfo}[2]{#2}
\providecommand{\eprint}[2][]{\url{#2}}

\bibitem[{\citenamefont{Porras and Cirac}(2006)}]{Porras2006b}
\bibinfo{author}{\bibfnamefont{D.}~\bibnamefont{Porras}} \bibnamefont{and}
  \bibinfo{author}{\bibfnamefont{J.~I.} \bibnamefont{Cirac}},
  \bibinfo{journal}{Phys. Rev. Lett.} \textbf{\bibinfo{volume}{96}},
  \bibinfo{pages}{250501} (\bibinfo{year}{2006}).

\bibitem[{\citenamefont{Blatt and Wineland}(2008)}]{Blatt2008}
\bibinfo{author}{\bibfnamefont{R.}~\bibnamefont{Blatt}} \bibnamefont{and}
  \bibinfo{author}{\bibfnamefont{D.}~\bibnamefont{Wineland}},
  \bibinfo{journal}{Nature} \textbf{\bibinfo{volume}{453}},
  \bibinfo{pages}{1008} (\bibinfo{year}{2008}).

\bibitem[{\citenamefont{Seidelin et~al.}(2006)\citenamefont{Seidelin,
  Chiaverini, Reichle, Bollinger, Leibfried, Britton, Wesenberg, Blakestad,
  Epstein, Hume et~al.}}]{Seidelin2006}
\bibinfo{author}{\bibfnamefont{S.}~\bibnamefont{Seidelin}},
  \bibinfo{author}{\bibfnamefont{J.}~\bibnamefont{Chiaverini}},
  \bibinfo{author}{\bibfnamefont{R.}~\bibnamefont{Reichle}},
  \bibinfo{author}{\bibfnamefont{J.~J.} \bibnamefont{Bollinger}},
  \bibinfo{author}{\bibfnamefont{D.}~\bibnamefont{Leibfried}},
  \bibinfo{author}{\bibfnamefont{J.}~\bibnamefont{Britton}},
  \bibinfo{author}{\bibfnamefont{J.~H.} \bibnamefont{Wesenberg}},
  \bibinfo{author}{\bibfnamefont{R.~B.} \bibnamefont{Blakestad}},
  \bibinfo{author}{\bibfnamefont{R.~J.} \bibnamefont{Epstein}},
  \bibinfo{author}{\bibfnamefont{D.~B.} \bibnamefont{Hume}},
  \bibnamefont{et~al.}, \bibinfo{journal}{Phys. Rev. Lett.}
  \textbf{\bibinfo{volume}{96}}, \bibinfo{pages}{253003}
  (\bibinfo{year}{2006}).

\bibitem[{\citenamefont{Amini et~al.}(2008)\citenamefont{Amini, Britton,
  Leibfried, and Wineland}}]{Amini2008}
\bibinfo{author}{\bibfnamefont{J.~M.} \bibnamefont{Amini}},
  \bibinfo{author}{\bibfnamefont{J.}~\bibnamefont{Britton}},
  \bibinfo{author}{\bibfnamefont{D.}~\bibnamefont{Leibfried}},
  \bibnamefont{and} \bibinfo{author}{\bibfnamefont{D.~J.}
  \bibnamefont{Wineland}}, \bibinfo{journal}{arXiv:0812.3907v1}
  (\bibinfo{year}{2008}).

\bibitem[{\citenamefont{Chiaverini and Lybarger}(2008)}]{Chiaverini2008}
\bibinfo{author}{\bibfnamefont{J.}~\bibnamefont{Chiaverini}} \bibnamefont{and}
  \bibinfo{author}{\bibfnamefont{W.~E.} \bibnamefont{Lybarger},
  \bibfnamefont{Jr.}}, \bibinfo{journal}{Phys. Rev. A}
  \textbf{\bibinfo{volume}{77}}, \bibinfo{pages}{022324}
  (\bibinfo{year}{2008}).

\bibitem[{\citenamefont{Ospelkaus et~al.}(2008)\citenamefont{Ospelkaus, Langer,
  Amini, Brown, Leibfried, and Wineland}}]{Ospelkaus2008}
\bibinfo{author}{\bibfnamefont{C.}~\bibnamefont{Ospelkaus}},
  \bibinfo{author}{\bibfnamefont{C.~E.} \bibnamefont{Langer}},
  \bibinfo{author}{\bibfnamefont{J.~M.} \bibnamefont{Amini}},
  \bibinfo{author}{\bibfnamefont{K.~R.} \bibnamefont{Brown}},
  \bibinfo{author}{\bibfnamefont{D.}~\bibnamefont{Leibfried}},
  \bibnamefont{and} \bibinfo{author}{\bibfnamefont{D.~J.}
  \bibnamefont{Wineland}}, \bibinfo{journal}{Phys. Rev. Lett.}
  \textbf{\bibinfo{volume}{101}}, \bibinfo{pages}{090502}
  (\bibinfo{year}{2008}).

\bibitem[{\citenamefont{Wineland et~al.}(1998)\citenamefont{Wineland, Monroe,
  Itano, Leibfried, King, and Meekhof}}]{Wineland1998}
\bibinfo{author}{\bibfnamefont{D.~J.} \bibnamefont{Wineland}},
  \bibinfo{author}{\bibfnamefont{C.}~\bibnamefont{Monroe}},
  \bibinfo{author}{\bibfnamefont{W.~M.} \bibnamefont{Itano}},
  \bibinfo{author}{\bibfnamefont{D.}~\bibnamefont{Leibfried}},
  \bibinfo{author}{\bibfnamefont{B.~E.} \bibnamefont{King}}, \bibnamefont{and}
  \bibinfo{author}{\bibfnamefont{D.~M.} \bibnamefont{Meekhof}},
  \bibinfo{journal}{J. Res. Natl. Inst. Stand. Technol.}
  \textbf{\bibinfo{volume}{103}}, \bibinfo{pages}{259} (\bibinfo{year}{1998}).

\bibitem[{\citenamefont{Turchette et~al.}(2000)\citenamefont{Turchette,
  Kielpinski, King, Leibfried, Meekhof, Myatt, Rowe, Sackett, Wood, Itano
  et~al.}}]{Turchette2000}
\bibinfo{author}{\bibfnamefont{Q.~A.} \bibnamefont{Turchette}},
  \bibinfo{author}{\bibfnamefont{D.}~\bibnamefont{Kielpinski}},
  \bibinfo{author}{\bibfnamefont{B.~E.} \bibnamefont{King}},
  \bibinfo{author}{\bibfnamefont{D.}~\bibnamefont{Leibfried}},
  \bibinfo{author}{\bibfnamefont{D.~M.} \bibnamefont{Meekhof}},
  \bibinfo{author}{\bibfnamefont{C.~J.} \bibnamefont{Myatt}},
  \bibinfo{author}{\bibfnamefont{M.~A.} \bibnamefont{Rowe}},
  \bibinfo{author}{\bibfnamefont{C.~A.} \bibnamefont{Sackett}},
  \bibinfo{author}{\bibfnamefont{C.~S.} \bibnamefont{Wood}},
  \bibinfo{author}{\bibfnamefont{W.~M.} \bibnamefont{Itano}},
  \bibnamefont{et~al.}, \bibinfo{journal}{Phys. Rev. A}
  \textbf{\bibinfo{volume}{61}}, \bibinfo{pages}{063418}
  (\bibinfo{year}{2000}).

\bibitem[{\citenamefont{Dehmelt}(1967)}]{Dehmelt1967}
\bibinfo{author}{\bibfnamefont{H.~G.} \bibnamefont{Dehmelt}},
  \bibinfo{journal}{Adv. At. Mol. Phys.} \textbf{\bibinfo{volume}{3}},
  \bibinfo{pages}{53} (\bibinfo{year}{1967}).

\bibitem[{\citenamefont{Leibfried et~al.}(2003)\citenamefont{Leibfried, Blatt,
  Monroe, and Wineland}}]{Leibfried2003}
\bibinfo{author}{\bibfnamefont{D.}~\bibnamefont{Leibfried}},
  \bibinfo{author}{\bibfnamefont{R.}~\bibnamefont{Blatt}},
  \bibinfo{author}{\bibfnamefont{C.}~\bibnamefont{Monroe}}, \bibnamefont{and}
  \bibinfo{author}{\bibfnamefont{D.}~\bibnamefont{Wineland}},
  \bibinfo{journal}{Rev. Mod. Phys.} \textbf{\bibinfo{volume}{75}},
  \bibinfo{pages}{281} (\bibinfo{year}{2003}).

\bibitem[{\citenamefont{House}(2008)}]{House2008}
\bibinfo{author}{\bibfnamefont{M.~G.} \bibnamefont{House}},
  \bibinfo{journal}{Phys. Rev. A} \textbf{\bibinfo{volume}{78}},
  \bibinfo{pages}{033402} (\bibinfo{year}{2008}).

\bibitem{footnote1} In Ref.~\cite{Wesenberg2008} an alternative definition $\kappa_j = q_z/q_0=| \partial_{z z}\ep(\vect{r}_j) | \times (z_j^2/U\si{rf})$ is used; the present algorithm is independent of the precise definition of $\kappa_j$.

\bibitem[{\citenamefont{Wesenberg}(2008)}]{Wesenberg2008}
\bibinfo{author}{\bibfnamefont{J.~H.} \bibnamefont{Wesenberg}},
  \bibinfo{journal}{Phys. Rev. A} \textbf{\bibinfo{volume}{78}},
  \bibinfo{pages}{063410} (\bibinfo{year}{2008}), \bibinfo{note}{erratum: after
  Eq.~(10) the maximum microtrap strength is $q_z\approx0.236q_0$.}

\bibitem[{\citenamefont{Papadimitriou}(1994)}]{Papadimitriou}
\bibinfo{author}{\bibfnamefont{C.~H.} \bibnamefont{Papadimitriou}},
  \emph{\bibinfo{title}{Computational Complexity}}
  (\bibinfo{publisher}{Addison-Wesley}, \bibinfo{year}{1994}).

\bibitem[{\citenamefont{Matou{\v s}ek and
  G{\"a}rtner}(2007)}]{MatousekGaertner}
\bibinfo{author}{\bibfnamefont{J.}~\bibnamefont{Matou{\v s}ek}}
  \bibnamefont{and}
  \bibinfo{author}{\bibfnamefont{B.}~\bibnamefont{G{\"a}rtner}},
  \emph{\bibinfo{title}{Understanding and Using Linear Programming}}
  (\bibinfo{publisher}{Springer}, \bibinfo{address}{Berlin Heidelberg},
  \bibinfo{year}{2007}).

\bibitem[{\citenamefont{Micheli et~al.}(2006)\citenamefont{Micheli, Brennen,
  and Zoller}}]{Micheli2006}
\bibinfo{author}{\bibfnamefont{A.}~\bibnamefont{Micheli}},
  \bibinfo{author}{\bibfnamefont{G.~K.} \bibnamefont{Brennen}},
  \bibnamefont{and} \bibinfo{author}{\bibfnamefont{P.}~\bibnamefont{Zoller}},
  \bibinfo{journal}{Nature Physics} \textbf{\bibinfo{volume}{2}},
  \bibinfo{pages}{341} (\bibinfo{year}{2006}).

\end{thebibliography}

\end{document}